
\input epsf
%\input youngtab
%!TEX TS-program = tex
%%%%%%%%%%%%%%%%%%%%%%%%%%%%%%%%%%%%%%%%%%%%%%%%%%%%%%%%%%%%%%%%%
%                                                               %
%       FONT FAMILIES:                                          %
%                                                               %
%%%%%%%%%%%%%%%%%%%%%%%%%%%%%%%%%%%%%%%%%%%%%%%%%%%%%%%%%%%%%%%%%
%                                                               %
%       Define script letters as rsfs                           %
%               (or redefine as cal)                            %
%                                                               %
%                                                               %
%%%%%%%%%%%%%%%%%%%%%%%%%%%%%%%%%%%%%%%%%%%%%%%%%%%%%%%%%%%%%%%%%
\newfam\scrfam
\batchmode\font\tenscr=rsfs10 \errorstopmode
\ifx\tenscr\nullfont
        \message{rsfs script font not available. Replacing with calligraphic.}
        
\else   
        \font\sevenscr=rsfs7
        \font\fivescr=rsfs5
        \skewchar\tenscr='177 \skewchar\sevenscr='177 \skewchar\fivescr='177
        \textfont\scrfam=\tenscr \scriptfont\scrfam=\sevenscr
        \scriptscriptfont\scrfam=\fivescr

\fi
%%%%%%%%%%%%%%%%%%%%%%%%%%%%%%%%%%%%%%%%%%%%%%%%%%%%%%%%%%%%%%%%%
%                                                               %
%       fraktur (or redefine as italic)		                %
%                                                               %
%%%%%%%%%%%%%%%%%%%%%%%%%%%%%%%%%%%%%%%%%%%%%%%%%%%%%%%%%%%%%%%%%
\catcode`\@=11
\newfam\frakfam
\batchmode\font\tenfrak=eufm10 \errorstopmode
\ifx\tenfrak\nullfont
        \message{eufm font not available. Replacing with italic.}
        \def\frak{\it}
\else
	
	\font\sevenfrak=eufm7 \font\fivefrak=eufm5
	\textfont\frakfam=\tenfrak
	\scriptfont\frakfam=\sevenfrak \scriptscriptfont\frakfam=\fivefrak
	\def\frak{\fam\frakfam}
\fi
\catcode`\@=\active
%%%%%%%%%%%%%%%%%%%%%%%%%%%%%%%%%%%%%%%%%%%%%%%%%%%%%%%%%%%%%%%%%
%                                                               %
%       Blackboard bold (or redefine as boldface)               %
%                                                               %
%%%%%%%%%%%%%%%%%%%%%%%%%%%%%%%%%%%%%%%%%%%%%%%%%%%%%%%%%%%%%%%%%
\newfam\msbfam
\batchmode\font\twelvemsb=msbm10 scaled\magstep1 \errorstopmode
\ifx\twelvemsb\nullfont\def\Bbb{\bf}

	\message{Blackboard bold not available. Replacing with boldface.}
\else   \catcode`\@=11
        \font\tenmsb=msbm10 \font\sevenmsb=msbm7 \font\fivemsb=msbm5
        \textfont\msbfam=\tenmsb
        \scriptfont\msbfam=\sevenmsb \scriptscriptfont\msbfam=\fivemsb
        \def\Bbb{\relax\expandafter\Bbb@}
        \def\Bbb@#1{{\Bbb@@{#1}}}
        \def\Bbb@@#1{\fam\msbfam\relax#1}
        \catcode`\@=\active

\fi
%%%%%%%%%%%%%%%%%%%%%%%%%%%%%%%%%%%%%%%%%%%%%%%%%%%%%%%%%%%%%%%%%
%                                                               %
%       More FONTS:                                             %
%                                                               %
%%%%%%%%%%%%%%%%%%%%%%%%%%%%%%%%%%%%%%%%%%%%%%%%%%%%%%%%%%%%%%%%%
        \font\eightrm=cmr8              \def\xrm{\eightrm}
        \font\eightbf=cmbx8             \def\xbf{\eightbf}
        \font\eightit=cmti10 at 8pt     \def\xit{\eightit}
%%%     \font\eightit=cmti8             \def\xit{\eightit}
                     
        \font\eightcp=cmcsc8
        \font\eighti=cmmi8              \def\xold{\eighti}
        \font\eightib=cmmib8             \def\xbold{\eightib}
        \font\teni=cmmi10               \def\old{\teni}
        \font\tencp=cmcsc10

        \font\twelvecp=cmcsc10 scaled\magstep1

        \font\twelvemath=cmmi12

\font\seventeenmath=cmmi10 at 17pt

	 at10pt	
	\font\twelvehelvbold=phvb at12pt
	 at14pt
	\font\sixteenhelvbold=phvb at16pt

\def\noblackbox{\overfullrule=0pt}
\noblackbox

%%%%%%%%%%%%%%%%%%%%%%%%%%%%%%%%%%%%%%%%%%%%%%%%%%%%%%%%%%%%%%%%%
%                                                               %
%       HEADLINE:                                               %
%                                                               %
%%%%%%%%%%%%%%%%%%%%%%%%%%%%%%%%%%%%%%%%%%%%%%%%%%%%%%%%%%%%%%%%%
\newtoks\headtext
\headline={\ifnum\pageno=1\hfill\else
	\ifodd\pageno{\eightcp\the\headtext}{ }\dotfill{ }{\old\folio}
	\else{\old\folio}{ }\dotfill{ }{\eightcp\the\headtext}\fi
	\fi}
\def\makeheadline{\vbox to 0pt{\vss\noindent\the\headline\break
\hbox to\hsize{\hfill}}
        \vskip2\baselineskip}
%%%%%%%%%%%%%%%%%%%%%%%%%%%%%%%%%%%%%%%%%%%%%%%%%%%%%%%%%%%%%%%%%
%                                                               %
%       FOOTNOTES:                                              %
%                                                               %
%%%%%%%%%%%%%%%%%%%%%%%%%%%%%%%%%%%%%%%%%%%%%%%%%%%%%%%%%%%%%%%%%
\newcount\infootnote
\infootnote=0
\def\foot#1#2{\infootnote=1
\footnote{${}^{#1}$}{\vtop{\baselineskip=.75\baselineskip
\advance\hsize by -\parindent\noindent{\xrm #2}}}\infootnote=0$\,$}
%%%%%%%%%%%%%%%%%%%%%%%%%%%%%%%%%%%%%%%%%%%%%%%%%%%%%%%%%%%%%%%%%
%                                                               %
%       REFERENCES:                                             %
%                                                               %
%%%%%%%%%%%%%%%%%%%%%%%%%%%%%%%%%%%%%%%%%%%%%%%%%%%%%%%%%%%%%%%%%
\newcount\refcount
\refcount=1
\newwrite\refwrite
\def\oldsize{\ifnum\infootnote=1\xold\else\old\fi}
\def\ref#1#2{
	\def#1{{{\oldsize\the\refcount}}\ifnum\the\refcount=1\immediate\openout\refwrite=\jobname.refs\fi\immediate\write\refwrite{\item{[{\xold\the\refcount}]} 
	#2\hfill\par\vskip-2pt}\xdef#1{{\noexpand\oldsize\the\refcount}}\global\advance\refcount by 1}
	}
\def\refout{\catcode`\@=11
        \xrm\immediate\closeout\refwrite
        \vskip2\baselineskip
        {\noindent\twelvecp References}\hfill\vskip\baselineskip
                                                %\vskip.25\baselineskip%%%%
        %\parskip=.875\parskip
        %\baselineskip=.8\baselineskip
        \baselineskip=.75\baselineskip
        \input\jobname.refs
        %\parskip=8\parskip \divide\parskip by 7
        %\baselineskip=1.25\baselineskip
        \baselineskip=4\baselineskip \divide\baselineskip by 3
        \catcode`\@=\active\rm}

\def\hepth#1{\href{http://arxiv.org/abs/hep-th/#1}{hep-th/{\xold#1}}}

\def\CQG#1#2#3{Class. Quantum Grav. {\xbold#1} ({\xold#2}) {\xold#3}}
\def\GRG#1#2#3{Gen. Rel. Grav. {\xbold#1} ({\xold#2}) {\xold#3}}

\def\PRD#1#2#3{Phys. Rev. {\xbf D}{\xbold#1} ({\xold#2}) {\xold#3}}

%%%%%%%%%%%%%%%%%%%%%%%%%%%%%%%%%%%%%%%%%%%%%%%%%%%%%%%%%%%%%%%%%
%                                                               %
%       SECTION NUMBERING:                                      %
%                                                               %
%%%%%%%%%%%%%%%%%%%%%%%%%%%%%%%%%%%%%%%%%%%%%%%%%%%%%%%%%%%%%%%%%
\newcount\sectioncount
\sectioncount=0
\def\section#1#2{\global\eqcount=0
	\global\subsectioncount=0
        \global\advance\sectioncount by 1
	\ifnum\sectioncount>1
	        \vskip2\baselineskip
	\fi
	\noindent
        \line{\twelvecp\the\sectioncount. #2\hfill}
		\vskip.8\baselineskip\noindent
        \xdef#1{{\old\the\sectioncount}}}
\newcount\subsectioncount
\def\subsection#1#2{\global\advance\subsectioncount by 1
	\vskip.8\baselineskip\noindent
	\line{\tencp\the\sectioncount.\the\subsectioncount. #2\hfill}
	\vskip.5\baselineskip\noindent
	\xdef#1{{\old\the\sectioncount}.{\old\the\subsectioncount}}}
\newcount\appendixcount
\appendixcount=0
\def\appendix#1{\global\eqcount=0
        \global\advance\appendixcount by 1
        \vskip2\baselineskip\noindent
        \ifnum\the\appendixcount=1
        \hbox{\twelvecp Appendix A: #1\hfill}\vskip\baselineskip\noindent\fi
    \ifnum\the\appendixcount=2
        \hbox{\twelvecp Appendix B: #1\hfill}\vskip\baselineskip\noindent\fi
    \ifnum\the\appendixcount=3
        \hbox{\twelvecp Appendix C: #1\hfill}\vskip\baselineskip\noindent\fi}
\def\oneappendix#1{\global\eqcount=0
        \global\advance\appendixcount by 1
        \vskip2\baselineskip\noindent
        \hbox{\twelvecp Appendix: #1\hfill}\vskip\baselineskip\noindent
    }
\def\acknowledgements{\vskip2\baselineskip\noindent
        \underbar{\it Acknowledgements:}\ }
%%%%%%%%%%%%%%%%%%%%%%%%%%%%%%%%%%%%%%%%%%%%%%%%%%%%%%%%%%%%%%%%%
%                                                               %
%       EQUATION NUMBERING                                      %
%                                                               %
%%%%%%%%%%%%%%%%%%%%%%%%%%%%%%%%%%%%%%%%%%%%%%%%%%%%%%%%%%%%%%%%%
\newcount\eqcount
\eqcount=0
\def\Eqn#1{\global\advance\eqcount by 1
\ifnum\the\sectioncount=0
	\xdef#1{{\old\the\eqcount}}
	\eqno({\oldstyle\the\eqcount})
\else
        \ifnum\the\appendixcount=0
	        \xdef#1{{\old\the\sectioncount}.{\old\the\eqcount}}
                \eqno({\oldstyle\the\sectioncount}.{\oldstyle\the\eqcount})\fi
        \ifnum\the\appendixcount=1
	        \xdef#1{{\oldstyle A}.{\old\the\eqcount}}
                \eqno({\oldstyle A}.{\oldstyle\the\eqcount})\fi
        \ifnum\the\appendixcount=2
	        \xdef#1{{\oldstyle B}.{\old\the\eqcount}}
                \eqno({\oldstyle B}.{\oldstyle\the\eqcount})\fi
        \ifnum\the\appendixcount=3
	        \xdef#1{{\oldstyle C}.{\old\the\eqcount}}
                \eqno({\oldstyle C}.{\oldstyle\the\eqcount})\fi
\fi}
\def\eqn{\global\advance\eqcount by 1
\ifnum\the\sectioncount=0
	\eqno({\oldstyle\the\eqcount})
\else
        \ifnum\the\appendixcount=0
                \eqno({\oldstyle\the\sectioncount}.{\oldstyle\the\eqcount})\fi
        \ifnum\the\appendixcount=1
                \eqno({\oldstyle A}.{\oldstyle\the\eqcount})\fi
        \ifnum\the\appendixcount=2
                \eqno({\oldstyle B}.{\oldstyle\the\eqcount})\fi
        \ifnum\the\appendixcount=3
                \eqno({\oldstyle C}.{\oldstyle\the\eqcount})\fi
\fi}
\def\multi{\global\advance\eqcount by 1}
\def\multieq#1#2{\xdef#1{{\old\the\eqcount#2}}
        \eqno{({\oldstyle\the\eqcount#2})}}
%%%%%%%%%%%%%%%%%%%%%%%%%%%%%%%%%%%%%%%%%%%%%%%%%%%%%%%%%%%%%%%%%
%                                                               %
%       Hyperrefs:                                          	%
%                                                               %
%%%%%%%%%%%%%%%%%%%%%%%%%%%%%%%%%%%%%%%%%%%%%%%%%%%%%%%%%%%%%%%%%
\newtoks\url
\def\Href#1#2{\catcode`\#=12\url={#1}\catcode`\#=\active#2}
\def\href#1#2{{#2}}

%%%%%%%%%%%%%%%%%%%%%%%%%%%%%%%%%%%%%%%%%%%%%%%%%%%%%%%%%%%%%%%%%
%                                                               %
%       FORMAT:                                                 %
%                                                               %
%%%%%%%%%%%%%%%%%%%%%%%%%%%%%%%%%%%%%%%%%%%%%%%%%%%%%%%%%%%%%%%%%
\parskip=3.5pt plus .3pt minus .3pt
\baselineskip=14pt plus .15pt minus .05pt
\lineskip=.5pt plus .05pt minus .05pt
\lineskiplimit=.5pt
\abovedisplayskip=18pt plus 4pt minus 2pt
\belowdisplayskip=\abovedisplayskip
\hsize=14cm
\vsize=19.5cm
\hoffset=1.5cm
\voffset=1.8cm
\frenchspacing
\footline={}
%\raggedbottom
%%%%%%%%%%%%%%%%%%%%%%%%%%%%%%%%%%%%%%%%%%%%%%%%%%%%%%%%%%%%%%%%%
%                                                               %
%       VARIOUS DEFINITIONS                                     %
%                                                               %
%%%%%%%%%%%%%%%%%%%%%%%%%%%%%%%%%%%%%%%%%%%%%%%%%%%%%%%%%%%%%%%%%

\def\ss{\scriptstyle}
\def\sss{\scriptscriptstyle}
\def\*{\partial}
\def\punkt{\,\,.}
\def\komma{\,\,,}

\def\={\!=\!}
\def\small#1{{\hbox{$#1$}}}

\def\fraction#1{\small{1\over#1}}
\def\fr{\fraction}
\def\Fraction#1#2{\small{#1\over#2}}
\def\Fr{\Fraction}
\def\tr{\hbox{\rm tr}}
\def\eg{{\tenit e.g.}}

\def\nlni{\hfill\break}

\def\a{\alpha}
\def\b{\beta}

\def\d{\delta}
\def\e{\varepsilon}

\def\G{\Gamma}

\def\Z{{\Bbb Z}}

%%%%%%%%%%%%%%%%%%%%%%%%%%%%%%%%%%%%%%%%%%%%%%%%%%%%%%%%%%%%%%%%%%%%%%%%%%%%%

%\leavevmode
%
%\special{!userdict begin /bop-hook{gsave 250 135 translate
%65 rotate /Times-Roman findfont 170 scalefont setfont
%0 0 moveto 0.7 setgray (DRAFT) show grestore}def end}

\headtext={Cederwall and Palmkvist: ``The octic $\ss E_8$ invariant''}

\line{
\epsfysize=20mm
\epsffile{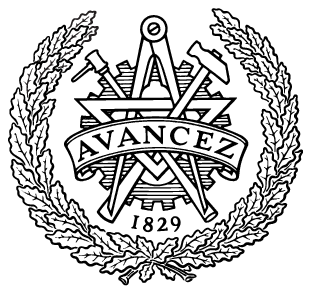}
\epsfysize=20mm
\epsffile{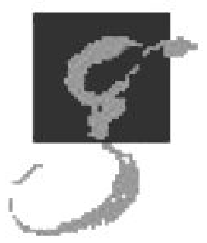}
\hfill}
\vskip-21mm
\line{\hfill G\"oteborg preprint}
\line{\hfill AEI-2007-005}
\line{\hfill hep-th/0702024}
\line{\hfill February, {\old2007}}
\line{\hrulefill}

\vfill

\centerline{\sixteenhelvbold The octic
{\seventeenmath E}\lower6pt\hbox{\twelvemath8} invariant}

\vfill

\centerline{\twelvehelvbold Martin Cederwall}

\vskip5mm

\centerline{\it Fundamental Physics}
\centerline{\it Chalmers University of Technology}
\centerline{\it S-412 96 G\"oteborg, Sweden}

\vskip15mm

\centerline{\twelvehelvbold Jakob Palmkvist}

\vskip5mm

\centerline{\it Max-Planck-Institut f\"ur Gravitationsphysik}
\centerline{\it Albert-Einstein-Institut}
\centerline{\it Am M\"uhlenberg 1, D-14476 Potsdam, Germany}

\vfill

{\narrower\noindent 
\underbar{Abstract:}
We give an explicit expression for the primitive $E_8$-invariant
tensor with eight symmetric indices. The result is presented in a manifestly 
$Spin(16)/\Z_2$-covariant notation.
\smallskip}
\vfill

\font\xxtt=cmtt6

\vtop{\baselineskip=.6\baselineskip\xxtt
\line{\hrulefill}
\catcode`\@=11
\line{email: martin.cederwall@chalmers.se, jakob.palmkvist@aei.mpg.de\hfill}
\catcode`\@=\active
}

\eject

\ref\FuchsSchweigert{J. Fuchs and C. Schweigert, 
{\xit ``Symmetries, Lie algebras and representations''},
Cambridge Univ. Press ({\xold1997}).}

\ref\LiE{A.M. Cohen, M. van Leeuwen and B. Lisser, 
LiE v. {\xold2}.{\xold2} ({\xold1998}), 
\nlni http://wallis.univ-poitiers.fr/\~{}maavl/LiE/} 

\ref\GAMMA{U. Gran,
{\xit ``GAMMA: A Mathematica package for performing gamma-matrix 
algebra and Fierz transformations in arbitrary dimensions''},
\hepth{0105086}.}

\ref\Dixmier{J. Dixmier, {\xit ``Enveloping algebras''}, 
North-Holland ({\xold1977}).}

\ref\Humphreys{J.E. Humphreys, {\xit ``Reflection groups and Coxeter
groups''}, Cambridge Univ. Press ({\xold1990}).}

\ref\Kac{V.G. Kac, {\xit ``Laplace operators of infinite-dimensional Lie
algebras and theta functions''}, 
Proc. Natl. Acad. Sci. USA, {\xbf81} ({\xold1984}) {\xold645}.}

\ref\Julia{B. Julia, {\xit ``Dualities in the classical supergravity limits''},
\hepth{9805083}.}

\ref\HigherDer{T. Damour and H. Nicolai, 
{\xit ``Higher order M theory corrections and the Kac--Moody 
algebra $\ss E_{\sss 10}$''}, \CQG{22}{2005}{2849} [\hepth{0504153}]; 
N. Lambert and P. West, 
{\xit ``Enhanced coset symmetries and higher derivative corrections''},
\PRD{74}{2006}{065002} [\hepth{0603255}]; 
T. Damour, A. Hanany, M. Henneaux, A. Kleinschmidt and H. Nicolai, 
{\xit ``Curvature corrections and Kac--Moody compatibility conditions''}, 
\GRG{38}{2006}{1507} [\hepth{0604143}].}

\def\f{\phi}

\def\ee{{\frak e}}
\def\ff{{\frak f}}
\def\gg{{\frak g}}
\def\hh{{\frak h}}
\def\kk{{\frak k}}
\def\so{{\frak so}}

\section\Introduction{Introduction and preliminaries}The largest of the
finite-dimensional exceptional Lie groups, $E_8$, with Lie algebra
$\ee_8$, is an interesting
object, both from a mathematical and a physical point of view.
It is an extraordinarily symmetric object, which \eg\ is reflected by the fact
that its root lattice is the unique even self-dual lattice in eight dimensions
(in euclidean space, even self-dual lattices only exist in dimension $8n$).
This property is essential for the existence of the $E_8\times E_8$ heterotic 
string. Because of self-duality, there is only one conjugacy class of
representations, the weight lattice equals the root lattice, and there is 
no ``fundamental'' representation smaller than the adjoint.
As one in the $E$-series of algebras, $E_8$ is relevant as a U-duality
group of symmetries for compactification of M-theory to three dimensions (see
\eg\ ref. [\Julia]).

In contrast to the large amount of elegance, calculations involving
$E_8$ and representations of $E_8$ are generically very complicated.
Anything resembling a tensor formalism is completely lacking.
A basic ingredient in a tensor calculus is a set of invariant tensors,
or ``Clebsch--Gordan coefficients''. The only invariant tensors
that are known explicitly for $E_8$ are the Killing metric and the
structure constants (which by definition take analogous forms for any
semi-simple Lie algebra in a Cartan--Weyl basis). The goal of this paper is
to take a first step towards a tensor formalism for $E_8$ by explicitly 
constructing an invariant tensor with eight symmetric adjoint indices.
The motivation for our work is partly mathematical and partly physical. 
On the mathematical side, the disturbing absence of a concrete 
expression for this tensor is unique among the 
finite-dimensional Lie groups. Even for the smaller exceptional algebras
$\gg_2$, $\ff_4$, $\ee_6$ and $\ee_7$, all invariant tensors are accessible
in explicit forms, due to the existence of ``fundamental'' representations 
smaller than the adjoint and to the connections with octonions and Jordan
algebras. On the physical side, we anticipate applications to U-duality 
in the presence of higher-derivative terms [\HigherDer].

The orders of Casimir invariants are known for all 
finite-dimensional semi-simple Lie algebras.
They are polynomials in $U(\gg)$, the universal 
enveloping algebra of $\gg$, of the form
$t_{A_1\ldots A_k}T^{A_1}\ldots T^{A_k}$, where $t$ is a symmetric invariant 
tensor and $T$ are generators of the algebra, and they generate the center 
$U(\gg)^\gg$ of $U(\gg)$.
The Harish-Chandra homomorphism
is the restriction of an element in $U(\gg)^\gg$ to a polynomial in 
the Cartan subalgebra $\hh$, which will be invariant under the 
Weyl group $W(\gg)$ of $\gg$. Due
to the fact that the Harish-Chandra homomorphism is an isomorphism 
from $U(\gg)^\gg$ to $U(\hh)^{W(\gg)}$
one may equivalently consider finding a basis of generators for the 
latter, a much easier problem. The orders of the invariants follow more or
less directly from a diagonalisation of the Coxeter element, the product 
of the simple Weyl reflections (see \eg\ refs. [\Humphreys,\Dixmier]). For 
infinite-dimensional algebras, one has to consider a completion of the
universal enveloping algebra in order to find invariants beyond the
Killing metric [\Kac]. 

In the case of $\ee_8$, the center $U(\ee_8)^{\ee_8}$ of the universal 
enveloping subalgebra is generated by elements of
orders 2, 8, 12, 14, 18, 20, 24 and 30.
The quadratic and octic invariants correspond to primitive invariant tensors
in terms of which the higher ones should be expressible. 
While the quadratic invariant is described by the Killing metric, 
the explicit form of the octic invariant
is previously not known (see ref. [\FuchsSchweigert], p. 304).
It is reasonable to assume that it will have an application in
the construction of higher-derivative deformations of M-theory compactified to
three (and lower) dimensions.

Lifting an element in $U(\hh)^{W(\gg)}$ back to $U(\gg)^\gg$ when $\gg$ is
not a matrix algebra and the corresponding invariant tensors are not known
may be a tedious problem. We would like to spend a moment considering 
the choice of method. An obvious path to follow is to consider manifest
symmetry only under a maximal subgroup $F\subset G=E_8$, and make an Ansatz for 
the invariant in terms of $F$-invariants. $E_8$ has a number of maximal 
subgroups, but one of them, $Spin(16)/\Z_2$, is natural for several reasons.
Considering calculational complexity, this is the subgroup that leads
to the smallest number of terms in the Ansatz. Considering the connection
to the Harish-Chandra homomorphism, $K=Spin(16)/\Z_2$ is the maximal compact 
subgroup of the split form $G=E_{8(8)}$. The Weyl group is a discrete subgroup
of $K$, and the Cartan subalgebra $\hh$ lies entirely in the coset directions 
$\gg/\kk$ (these statements apply in general). Finally, considering physical
applications, $G/K$ cosets, with $K$ the maximal compact subgroup of the 
split form of $G$, are the ones occurring in sigma models for M-theory
compactifications.

There is indeed a significant intermediary step in the Harish-Chandra
homomorphism. 
Consider it as the decomposition $f\circ e$ of $e$: 
$U(\gg)^\gg\rightarrow U(\gg/\kk)^\kk$ and 
$f$: $U(\gg/\kk)^\kk\rightarrow U(\hh)^{W(\gg)}$, where both $e$ and $f$
act as restrictions (the notation $U(\gg/\kk)$ is of course not to be
interpreted in the sense of a universal enveloping algebra, it is the
space of polynomials on $\gg/\kk$). 
The operator $e$ obviously exists for any subalgebra,
not only $\kk$, and $f$ exists thanks to $\hh\subset\gg/\kk$ and 
$W(\gg)\subset K$. Since the Harish-Chandra homomorphism is an isomorphism, 
both $e$ and $f$ are isomorphisms as well.
This is relevant for 
higher-derivative terms in sigma model actions, which (modulo multiplication
by automorphic forms) then can equivalently
be written as terms in $U(\gg)^\gg$ with a $\kk$-valued
Lagrange multiplier gauge connection, 
or as terms in $U(\gg/\kk)^\kk$ with the gauge connection eliminated.
One conceivable approach to finding the full invariant would be to 
start from $U(\ee_8/\so(16))^{\so(16)}$ and use an 
$E_8$ group element formed by exponentiating the spinor generators to
conjugate the spinor out in the full algebra. Our impression is that such 
a calculation would be at least as difficult as the direct check of invariance
performed below.

%\vfill\eject

\section\Calculation{The invariant}We thus consider the decomposition 
of the adjoint representation of $E_8$
into representations of the maximal subgroup $Spin(16)/\Z_2$. The adjoint
decomposes into the adjoint {\bf 120} and a chiral spinor {\bf
128}. We often use Dynkin labels for highest weights to label
representations; these have labels (01000000) and (00000010), respectively.
Our convention for chirality is $\G_{a_1\ldots a_{16}}\f=+\e_{a_1\ldots a_{16}}\f$.
The $\ee_8$ algebra becomes
$$
\eqalign{
[T^{ab},T^{cd}]&=2\d^{[a}_{[c}T_{\mathstrut}^{b]}{}^{\mathstrut}_{d]}\komma\cr
[T^{ab},\f^\a]&=\fr4(\G^{ab}\f)^\a\komma\cr
[\f^\a,\f^\b]&=\fr8(\G_{ab})^{\a\b}T^{ab}\punkt\cr
}\Eqn\Algebra
$$
The coefficients in the first and second commutators are related by the
$\so(16)$ algebra. The normalisation of the last commutator is free, but
is fixed by the choice for the quadratic invariant, which for the case
above is $X_2=\fr2T_{ab}T^{ab}+\f_\a\f^\a$. 
Spinor and vector indices are raised and lowered with $\d$.
Equation (\Algebra) describes the compact real form, $E_{8(-248)}$. By letting
$\f\rightarrow i\f$ one gets $E_{8(8)}$, where the spinor generators
are non-compact, which is the real form relevant as duality symmetry
in three dimensions (other real forms contain a non-compact $Spin(16)/\Z_2$
subgroup). 
The Jacobi identities are satisfied thanks to the Fierz identity
$(\G_{ab})_{[\a\b}(\G^{ab})_{\a\b]}=0$, which is satisfied for
$\so(8)$ with chiral spinors, 
$\so(9)$, and $\so(16)$ with chiral spinors 
(in the former cases the algebras are
$\so(9)$, due to triality, and ${\ff}_4$).

The Harish-Chandra homomorphism tells us that the ``heart'' of the invariant 
lies in an octic Weyl-invariant of the Cartan subalgebra. A first step may be
to lift it to a unique $Spin(16)/\Z_2$-invariant in the spinor, corresponding
to applying the isomorphism $f^{-1}$ above. It is gratifying to verify (using
\eg\ LiE [\LiE]) that there is indeed an octic invariant (other than
$(\f\f)^4$), and that no such invariant exists at lower order.
Using Fierz identities (more below and in the appendix), it is 
straightforward to show that the new invariant is proportional to
$$
\eqalign{
&(\f\G_{ab}{}^{cd}\f)(\f\G_{cd}{}^{ef}\f)
(\f\G_{ef}{}^{gh}\f)(\f\G_{gh}{}^{ab}\f)\cr
\hbox{or}\quad&
\e^{a_1\ldots a_{16}}(\f\G_{a_1a_2a_3a_4}\f)(\f\G_{a_5a_6a_7a_8}\f)
(\f\G_{a_9a_{10}a_{11}a_{12}}\f)(\f\G_{a_{13}a_{14}a_{15}a_{16}}\f)\cr
}\Eqn\Phiinvariant
$$
(the two expressions are proportional modulo $(\f\f)^4$).
This is the expression that would go into a deformation of the sigma model
without an $\so(16)$ gauge field.
Making an Ansatz for the entire $E_8$ invariant, we need to include also 
the generators $T^{ab}$, and write down the most general $\so(16)$-invariant
with terms of orders $T^8$, $T^6\f^2$, $T^4\f^4$, $T^2\f^6$, $\f^8$. The
number of these are 6, 11, 12, 5 and 2, respectively. 
We then have to check invariance
only under the action of the spinorial generators. Out of the 36 coefficients
in the general Ansatz, we expect 34 to become determined in terms of the
remaining two, giving a linear combination of the fourth power of the
quadratic invariant and a traceless octic invariant.

The counting can be refined to determine the coinciding irreducible
$\so(16)$ representations in $T^{8-2k}$ and $\f^{2k}$. 
This will give us a concrete
guideline in writing down the Ansatz. At this stage, exact Fierz identities 
are not needed, just the knowledge that they exist to make the Ansatz 
complete. Let us take some examples. At order $T^8$, the 6 independent terms 
are $\hbox{Pf}(T)$, $\tr T^8$, $\tr T^6\tr T^2$, $(\tr T^4)^2$, 
$\tr T^4(\tr T^2)^2$ and $(\tr T^2)^4$. At order $T^6\f^2$, $\f^2$
contains the representations
$\otimes_s^2(00000010)=(00000000)\oplus(00000020)\oplus(00010000)$
(see the appendix for Fierz identities).
These have to be contracted to singlets with the same representations in
$\otimes_s^6(01000000)
=3(00000000)\oplus2(00000020)\oplus6(00010000)\oplus\ldots\,\,$.
How these considerations go into the Ansatz is easily read off from the
final expression for the invariant below.
Going to higher order in $\f$ and lower in $T$ makes things more involved, 
although all one really has to take care of is to choose a linearly independent
set of expressions in $\phi$ when a representation occurs with multiplicity 
greater than one. We should mention that we do not actually work with 
irreducible representations. Forming an element of an irreducible representation
containing a number of spinors involves symmetrisations and
subtraction of traces, 
which can be rather complicated. This becomes even more pronounced when we are
dealing with transformation of terms in our Ansatz under the spinor generators,
which will transform as spinors. Then irreducibility also
involves gamma-trace conditions.
Instead we use simple 
expressions that we know contain the irreducible ones. To take an example 
at order $\f^4$, just considering the structure of the vector indices in
the expression $(\f\G_{ab}{}^{ij}\f)(\f\G_{cdij}\f)$ tells us that it may
contain the representations (00000000), (00010000), (02000000) and (20000000).
However, $\f^4$ contains no (02000000) and only one (00010000), 
which means that (02000000) vanishes and (00010000) can be 
represented by a ``simpler''
expression, $(\f\f)(\f\G_{abcd}\f)$, as we will see in the appendix. 
The (00000000) represents a trace that
we do not subtract explicitly. We simply use the above expression to ensure
that the representation (02000000) is present. 
So, our expressions in the Ansatz, and also in
the equations, which we will not display in detail, will be related to 
irreducible representations by a (block-)triangular matrix. 
The results of all these considerations can be read off from the resulting
invariant below.

The transformation of the Ansatz under the action of the spinorial
generator is an $\so(16)$ spinor. The vanishing of this spinor is 
equivalent to $\ee_8$ invariance. 
The spinorial generator acts similarly to a supersymmetry generator on
a superfield, giving terms at order $T^{7-2k}\f^{1+2k}$ from $T^{8-2k}\f^{2k}$ 
and $T^{6-2k}\f^{2+2k}$. Here, it is necessary to use the full machinery of
Fierz identities, some of which are described in the appendix. Even though
all identities may be derived from the ones at $\f^3$, it becomes increasingly 
difficult to do so by hand as the number of $\phi$'s increases. We have used
a combination of manual calculation and calculations in the Mathematica 
package GAMMA [\GAMMA], based on representation contents obtained with LiE
[\LiE]. The number of equations is the number of spinors that can be
formed as $T^{7-2k}\f^{1+2k}$. For $k=0,1,2,3$ these numbers 
are 15, 37, 25 and 5, 
respectively. It is satisfying to see that of these 82 equations, only 34
are linearly independent, as anticipated above. In addition, the result
is consistent with the form of the quadratic invariant, and it is possible
to form a traceless octic invariant tensor. All of this is obvious seen
from an $E_8$ perspective, but acts as a (much needed) consistency check 
on the calculations. None of the coefficients in the Ansatz is determined
by a single relation, most by several, so we are quite confident that our
result is correct.

The final result for the octic invariant is, up to an overall 
multiplicative constant:
$$
\eqalign{
X_8&=\fr{3072}\e^{a_1\ldots a_{16}}T_{a_1a_2}\ldots T_{a_{15}a_{16}}\cr
&\qquad-30\tr T^8+14\tr T^6\tr T^2+\Fr{35}4(\tr T^4)^2%\cr
%&\qquad
-\Fr{35}8\tr T^4(\tr T^2)^2+\Fr{15}{64}(\tr T^2)^4\cr
&+\bigl[2\tr T^6-\tr T^4\tr T^2+\fr8(\tr T^2)^3](\f\f)\cr
&\qquad+\bigl[\bigl(\Fr54\tr T^4-\fr2(\tr T^2)^2\bigr)T^{ab}T^{cd}
              +\Fr{27}4\tr T^2T^{ab}(T^3)^{cd} \cr
&\qquad\qquad-15T^{ab}(T^5)^{cd}-15(T^3)^{ab}(T^3)^{cd}\bigr]
      (\f\G_{abcd}\f)\cr
&\qquad+\bigl[\fr{16}\tr T^2T^{ab}T^{cd}T^{ef}T^{gh}
      -\Fr58T^{ab}T^{cd}T^{ef}(T^3)^{gh}\bigr]
      (\f\G_{abcdefgh}\f)\cr
&\qquad-\fr{192}T^{ab}T^{cd}T^{ef}T^{gh}T^{ij}T^{kl}(\f\G_{abcdefghijkl}\f)\cr
&+\bigl[7\tr T^4-\Fr{31}8(\tr T^2)^2\bigr](\f\f)^2\cr
&\qquad-\Fr3{64}T^{ab}T^{cd}T^{ef}T^{gh}(\f\f)(\f\G_{abcdefgh}\f)\cr
&\qquad+\bigl[\Fr5{64}T^{ab}T^{cd}T^{ef}T^{gh}
        -\Fr{15}{16}T^{ab}T^{ce}T^{df}T^{gh}\cr
&\qquad\qquad    +\Fr58T^{ae}T^{bf}T^{cg}T^{dh}\bigr]
        (\f\G_{abcd}\f)(\f\G_{efgh}\f)\cr
&\qquad+\bigl[\Fr32(T^3)^{ab}T^{cd}-\fr8\tr T^2T^{ab}T^{cd}\bigr]
          (\f\f)(\f\G_{abcd}\f)\cr
&\qquad+\bigl[\Fr{15}{16}(T^3)^{ab}T^{cd}-\Fr3{16}\tr T^2T^{ab}T^{cd}
             +\Fr54(T^2)^{ac}(T^2)^{bd}\bigr]
               (\f\G_{ab}{}^{ij}\f)(\f\G_{cdij}\f)\cr
&\qquad+\Fr{15}8T^{ab}T^{cd}(T^2)^{ef}(\f\G_{abe}{}^i\f)(\f\G_{cdfi}\f)\cr
&+\fr2\tr T^2(\f\f)^3+\Fr{55}{32}T^{ab}T^{cd}(\f\f)^2(\f\G_{abcd}\f)\cr
&\qquad+\fr8T^{ab}T^{cd}(\f\f)(\f\G_{ab}{}^{ij}\f)(\f\G_{cdij}\f)\cr
&\qquad+\bigl[-\fr{384}T^{ab}T^{cd}+\Fr7{192}T^{ac}T^{bd}\bigr]
        (\f\G_{ab}{}^{ij}\f)(\f\G_{cd}{}^{kl}\f)(\f\G_{ijkl}\f)\cr
&-\Fr{57}{32}(\f\f)^4%\cr
%&\qquad
+\fr{12288}(\f\G_{ab}{}^{cd}\f)(\f\G_{cd}{}^{ef}\f)
        (\f\G_{ef}{}^{gh}\f)(\f\G_{gh}{}^{ab}\f)\cr
&+\b[-\fr2\tr T^2+(\f\f)]^4\punkt\cr
}\Eqn\MainResult
$$
Here, $\b$ is an arbitrary constant multiplying the fourth power of
the quadratic invariant.
The trace vanishes for $\b=\Fr9{127}$ (that such a value exists at all is 
non-trivial and provides a further check on the coefficients).
The occurrence of the prime 127 is not incidental; taking the trace of 
$\d^{(AB}\d^{CD}\d^{EF}\d^{GH)}$ gives 
$\d_{GH}\d^{(AB}\d^{CD}\d^{EF}\d^{GH)}=(\fr7\cdot248+\Fr67)\d^{(AB}\d^{CD}\d^{EF)}
=\Fr{2\cdot127}7\d^{(AB}\d^{CD}\d^{EF)}$.
The actual technique we use for calculating the trace is not to extract
the eight-index tensor, but to act on the invariant with
$\fr2{\partial\over\partial T_{ab}}{\partial\over\partial T^{ab}}
+{\partial\over\partial\f_\a}{\partial\over\partial\f^\a}$.
We remind that eq. (\MainResult) gives the octic invariant for the compact form
$E_{8(-248)}$. The corresponding expression for the split form $E_{8(8)}$ is 
obtained by a sign change of the terms containing $\f^{4k+2}$.

It would of course be of great use if one could extend the present
investigation to a tensor formalism for $E_8$. Part of that project would
be to identify all relations that the octic invariant tensor fulfills. 
For example, there is no new invariant at order 10. This means that
$t^{(A_1\ldots A_5}{}_{BCD}t^{A_6\ldots A_{10})BCD}
=at^{(A_1\ldots A_8}\d^{A_9A_{10})}+b\d^{(A_1A_2}\ldots\d^{A_9A_{10})}$.
%$$
%t^{(A_1\ldots A_5}{}_{BCD}t^{A_6\ldots A_{10})BCD}
%=at^{(A_1\ldots A_8}\d^{A_9A_{10})}
%+b\d^{(A_1A_2}\ldots\d^{A_9A_{10})}
%\punkt\eqn
%$$
It is not within our power to check this to all orders. We have checked
 (using Mathematica, due to the complexity of differentiating and
tracing the expressions in $T$) 
that it, 
quite non-trivially from an $\so(16)$ perspective, happens at
lowest order in $\f$, $T^{10}$. When $t$ is traceless the coefficients are
$a=3\cdot5\cdot13\cdot23\cdot29/(2\cdot7\cdot127)$, 
$b=3^3\cdot5\cdot13\cdot19\cdot37/(2^2\cdot127^2)$.
We then have expressions also for the invariants at orders 12 and 14,
namely $t^{(A_1\ldots A_6}{}_{BC}t^{A_7\ldots A_{12})BC}$
and $t^{(A_1\ldots A_7}{}_Bt^{A_8\ldots A_{14})B}$.
In order to form higher invariants, one will need expressions with
more than two $t$'s.

In conclusion, it is satisfactory that the octic invariant can be constructed.
What one really would like to use it for is to derive identities for it, so
that its explicit form in some basis can be dropped. In the present 
framework this task looks very difficult, unless one may find a way of
automating the calculations. A possible refinement of the present
formalism, inspired by the
Harish-Chandra homomorphism, would be to derive higher Fierz identities 
in a specific $Spin(16)/\Z_2$ frame, where the spinor lies entirely
in the Cartan subalgebra. Such a formalism would presumably 
be straightforward to implement in Mathematica, and would lead to much less 
time-consuming calculations.

\oneappendix{Fierz identities}In this appendix we will 
describe the Fierz identities that we have used to find the linear 
dependence between 
representations at order $\f^{n}$ for $n \geq 3$. 
As explained above, for even $n$ we only have to make sure that the basis 
elements in the Ansatz are linearly independent, while for odd $n$, we 
need to know the exact dependence for determining the equations that the 
coefficients in the Ansatz must satisfy. 

Fierz identities relate different expressions with the same 
index structure. Our strategy is to go from  
lower to higher powers of $\phi$ and, for each order, 
with increasing number of indices. In this way, higher identities can be 
derived by hand from the lower ones, but the calculations will also 
generically be more and more complicated. Some of the identities have 
instead been obtained directly using the Mathematica package GAMMA.
We will not write down all the Fierz identities here, but explain 
the method with some examples, starting from the bottom. The symmetric 
product of two spinor representations decomposes into irreducible 
representations as
$$
\otimes^2_s(00000010)=(00000000)\oplus(00010000)\oplus(00000020)
\komma\eqn
$$
where the terms on the right hand side correspond to the basis elements 
$\f\f,\, \f\G_{abcd}\f$ and $\f\G_{abcdefgh}\f$
at order $\f^2$. We do not need any Fierz identities here, but proceed 
to $\phi^3$ where we have
$$
\eqalign{
\otimes^3_s(00000010)=(00000010)\oplus(01000010)\oplus(00010010)
\oplus(00000030)\punkt
}\eqn
$$
Again all the irreducible representations come with multiplicity one.
The corresponding terms will have one spinor index each, and 0, 2, 4 and 8 
antisymmetric vector indices, respectively. From the basis elements at 
$\phi^2$ we can construct three expressions with no free vector indices by 
multiplying by a spinor and an antisymmetric product of gamma matrices. 
Since there is only one $(00000010)$, any two of them must be linearly 
dependent and we make the Ansatz
$$
\eqalign{
\f(\f\f)=\Fr{A}{4!} \G^{abcd}\f(\f\G_{abcd}\f) = 
\Fr{B}{8!} \G^{abcdefgh}\f(\f\G_{abcdefgh}\f)\komma
}\eqn
$$
or equivalently, writing out the spinor indices,
$$
\eqalign{
\delta_{\alpha (\beta}\delta_{\gamma \delta)}
=\Fr{A}{4!} (\G^{abcd})_{\alpha (\beta}(\G_{abcd})_{\gamma \delta)} = 
\Fr{B}{8!} (\G^{abcdefgh})_{\alpha (\beta}(\G_{abcdefgh})_{\gamma \delta)}\punkt
}\eqn
$$
Contracting the equations with $\delta^{\beta \gamma}$ 
gives $A=\fr{28}$ and $B=\fr{198}$. 
We choose $\f(\f\f)$ as a basis element corresponding to (00000010), 
but we could of course also choose $\G^{abcd}\f(\f\G_{abcd}\f)$ 
or $\G^{abcdefgh}\f(\f\G_{abcdefgh}\f)$.
This method of contracting the spinor indices in an Ansatz to determine the 
coefficients is the one that we have implemented in GAMMA for direct 
calculations. In this example it is easily done by hand, but for more terms 
we have to contract not only with $\delta$, but also with $\G^{ijkl}$ or 
$\G^{ijklmnpq}$, and for higher orders in $\phi$ we have to perform multiple 
contractions, since each contraction removes two spinor indices. 
This complicate the calculations considerably, but the principle is the same.

We return to the representations at order $\f^3$. For expressions with 
free vector indices, we have to take into account that the irreducible 
representations are all gamma-traceless. 
This means that, in order to obtain (01000010), we must combine any of 
the two expressions 
$\G^{cd}\f(\f\G_{abcd}\f)$ and $\G^{cdefgh}\f(\f\G_{abcdefgh}\f)$, 
constructed from the basis elements at $\f^2$, with $\G^{ab}\f(\f\f)$, 
from the one that we already have at $\f^3$. 
However, we do not need these gamma-traceless linear combinations, 
only the relation between them. Since $(01000010)$ occurs with 
multiplicity one, they must be proportional to each other, which means 
that the three expressions 
$\G^{cd}\f(\f\G_{abcd}\f),\,\G^{cdefgh}\f(\f\G_{abcdefgh}\f)$ and 
$\G^{ab}\f(\f\f)$ are linearly dependent. This Ansatz leads to the 
Fierz identity
$$
\fr{6!}\G^{cdefgh}\f(\f\G_{abcdefgh}\f)=
-\G^{cd}\f(\f\G_{abcd}\f)-49\G_{ab}\f(\f\f)
\komma\eqn
$$
and we choose $\G^{cd}\f(\f\G_{abcd}\f)$ as a new basis element. 
This will in the same way give rise to terms corresponding to 
gamma-traces in Ans\"atze for expressions with more than two indices.
Since there is no (10000001) or (00100001), the expressions with one or 
three antisymmetric indices are pure gamma-traces. We write only two of the 
four identities here:
$$
\eqalign{
\G^{bcd}\f(\f\G_{abcd}\f)&=42\G_a\f(\f\f)\komma\cr
\G^d\f(\f\G_{abcd}\f)&=\fr4\G_{[a}\G^{de}\f(\f\G_{bc]de}\f)
+\fr2\G_{abc}\f(\f\f)\punkt
}\Eqn\trefierz
$$
Both of them can be used to obtain Fierz identities at $\f^4$.
The first one (multiplied by a gamma matrix and a spinor) shows that the 
(20000000) part of $(\f{\G_{ab}}^{ij}\f)(\f{\G_{cdij}}\f)$ vanishes. 
The second one is very useful for deriving higher Fierz identities in general. 
For example, we can apply it to the (00010000) part of the same expression,
$$
\eqalign{
&(\f{\G_{[ab}}^{ij}\f)(\f{\G_{cd]ij}}\f)
=(\f{\G_{[ab}}^{i}\G^{j}\f)(\f{\G_{cd]ij}}\f)\cr
&\qquad=\fr{12}(\f{\G_{[ab}}^{i}\G_{c}\G^{kl}\f)(\f\G_{d]ikl}\f)
+\fr{6}(\f{\G_{[ab}}^{i}\G_{|i|}\G^{kl}\f)(\f\G_{cd]kl}\f)\cr
&\qquad\qquad+\fr2(\f{\G_{[ab}}^{i}{\G_{cd]i}}\f)(\f\f)\cr
&\qquad=2(\f\G_{[ab}{}^{kl}\f)(\f\G_{cd]kl}\f)+4(\f\G_{abcd}\f)(\f\f)
\komma}\eqn
$$
%Evaluating the products of gamma matrices, this simplifies to
%$$
%\eqalign{
%2(\f\G_{[ab}{}^{kl}\f)(\f\G_{cd]kl}\f)+4(\f\G_{abcd}\f)(\f\f)
%}\eqn
%$$
and we see that the (00010000) at $\f^4$ is indeed represented by the 
``simpler'' 
expression $(\f\f)(\f\G_{abcd}\f)$.
%$$
%\eqalign{
%(\f\G_{[ab}{}^{ij}\f)(\f\G_{cd]ij}\f)=-4(\f\f)(\f\G_{abcd}\f)\punkt
%}\eqn
%$$

We end with an example of a $\phi^5$ identity, with a degree of
complexity which is typical for the ones we use at this level, 
obtained by means of GAMMA.
The identity, which relates seemingly different expressions for 
$(02000010)\oplus2(00010000)$, reads
$$
\eqalign{
\G^{ijkl}\f(\f\G_{abij}\f)&(\f\G_{cdkl}\f)
=-10\f(\f\G_{ab}{}^{ij}\f)(\f\G_{cdij}\f)+24\f(\f\f)(\f\G_{abcd}\f)\cr
&\quad-4\G^{ij}\f(\f\G_{[abc}{}^k\f)(\f\G_{d]ijk}\f)\cr
&\quad-6\G_{[a}\G^i\f(\f\G_{b]ijk}\f)(\f\G_{cd}{}^{jk}\f)
        -6\G_{[c}\G^i\f(\f\G_{d]ijk}\f)(\f\G_{ab}{}^{jk}\f)\cr
&\quad+\G_{ab}\G^{ij}\f(\f\G_{cdij}\f)(\f\f)
        +\G_{cd}\G^{ij}\f(\f\G_{abij}\f)(\f\f)\cr
&\quad+4\G_{[a|[c}\G^{ij}\f(\f\G_{d]|b]ij}\f)(\f\f)\cr
&\quad+\fr2\G_{ab}\G^{ij}\f(\f\G_{cd}{}^{kl}\f)(\f\G_{ijkl}\f)
        +\fr2\G_{cd}\G^{ij}\f(\f\G_{ab}{}^{kl}\f)(\f\G_{ijkl}\f)\cr
&\quad+12\G_{abcd}\f(\f\f)^2\punkt\cr
}\Eqn\FiveFierz
$$
Here, everything except the first three terms on the right hand side 
represents gamma-traces, whose exact form and coefficients are still important.
They are deduced from the representation content in $\oplus_s^5(00000010)$
with fewer than four vector indices, namely (11000001) (line 3), 2(01000010)
(lines 4-6) and (00000010) (line 7) (the full Ansatz contains another two
terms with (01000010) and (00000010), whose coefficients turn out to vanish). 
Rather than tracing four spinor indices in an Ansatz with these
terms, already containing free vector indices, with products of
symmetric elements in the Clifford algebra, we choose to form scalars
by tracing and contracting with a suitable
number of elements with the same tensor structure as the terms themselves. This
method turns out to be much less time-consuming.
It may seem that eq. (\FiveFierz) gives rise to a $\f^6$ identity for 
$(\f\G^{ijkl}\f)(\f\G_{abij}\f)(\f\G_{cdkl}\f)$ by multiplying by a spinor, 
which would then make our basis at $\f^6$ incomplete, but fortunately, 
this expression will be cancelled by terms on the right hand side.

\acknowledgements
We would like to thank Bengt EW Nilsson, Christoffer Petersson, 
Daniel Persson and Axel Kleinschmidt for discussions, 
and especially Ulf Gran, whose help 
with GAMMA, with Mathematica programming, with incorporating chirality 
in GAMMA and letting us
use a pre-release version has been essential for the completion
of our calculations.

%\vfill\eject

\refout

\end